# Spectral dynamics of topological shift-current in ferroelectric semiconductor SbSI

M. Sotome[1]*, M. Nakamura[1,2], J. Fujioka[2,3], M. Ogino[3], Y. Kaneko[1], T. Morimoto[4], Y. Zhang[4,5], M. Kawasaki[1,3], N. Nagaosa[1,3], Y. Tokura[1,3] and N. Ogawa[1,2]

[1]RIKEN Center for Emergent Matter Science (CEMS), Wako 351-0198, Japan.

[2]PRESTO, Japan Science and Technology Agency (JST), Kawaguchi 332-0012, Japan.

[3]Department of Applied Physics and Quantum Phase Electronics Center (QPEC), University of Tokyo, Tokyo 113-8656, Japan.

[4]Max Planck Institute for Chemical Physics of Solids, 01187 Dresden, Germany

[5]Leibniz Institute for Solid State and Materials Research, 01069 Dresden, Germany

Photoexcitation in solids brings about transitions of electrons/holes between different electronic bands. If the solid lacks an inversion symmetry, these electronic transitions support spontaneous photocurrent due to the topological character of the constituting electronic bands; the Berry connection. This photocurrent, termed shift current, is expected to emerge on the time-scale of primary photoexcitation process. We observed ultrafast time evolution of the shift current in a prototypical ferroelectric semiconductor by detecting emitted terahertz electromagnetic waves. By sweeping the excitation photon energy across the band gap, ultrafast electron dynamics as a source of terahertz emission abruptly changes its nature, reflecting a contribution of Berry connection upon interband optical transition. The shift excitation carries a net charge flow, and is followed by a swing-over of the electron cloud on the sub-picosecond time-scale of electron-phonon interaction. Understanding these substantive characters of the shift current will pave the way for its application to ultrafast sensors and solar cells.

**I. INTRODUCTION**

In modern theory of polarization, the spontaneous electronic polarization in a crystal is formulated as the quantum phenomenon in terms of the Berry phase [1,2]. The spontaneous photocurrent in the polar crystals has also been reviewed recently both in theory and experiment; the shift current (or shift photocurrent) due to the coherent shift of electron cloud in real space is found to emerge from a difference in the Berry connection between valence and conduction bands;

$$j = \frac{\pi e^2 |E_{\text{exc}}|^2}{\hbar^2 \omega_{\text{exc}}^2} \int d\mathbf{k}\, \delta(\epsilon_1 - \epsilon_2 + \hbar\omega_{\text{exc}}) |v_{12}|^2 R_{\mathbf{k}}, \quad (1)$$

$$R_{\mathbf{k}} = \frac{\partial \varphi_{12}}{\partial \mathbf{k}} + \mathbf{a}_1 - \mathbf{a}_2, \quad (2)$$

where $e$ is the elementary charge, $E_{\text{exc}}$ and $\omega_{\text{exc}}$ the electric field and angular frequency of light, respectively, $v_{12}$ the transition matrix element, $\epsilon_1$ ($\epsilon_2$) the energy of the valence (conduction) band, $\hbar$ the Planck constant divided by $2\pi$. Here $R_k$ is the gauge-independent shift vector, with $\varphi_{12}$ the phase of $v_{12}$, and $a_1$ ($a_2$) the Berry connection $-i<u_{n\mathbf{k}}|\nabla_{\mathbf{k}} u_{n\mathbf{k}}>$ of the valence (conduction) band [3]. Historically, the shift current has been discussed as anomalous bulk photovoltaic/galvanic effects [4,5], and described in several ways including second-order nonlinear optics [6], kinetic theory [7], and Floquet formalism [8]. First-principles calculations with the above equations have successfully reproduced the experimentally observed spectral features of the shift current in actual ferroelectric materials [3]. These quantum theories have clearly unveiled the topological nature of the shift current, and the concept of the topological shift current is now extended to the cases of excitonic process [9] and spin current [10]. Experimentally, the shift current has been demonstrated in various systems [11-15], and is proposed to be a major factor for realizing highly-efficient solar cells [16-18]. Since the shift current initiates upon an interband photoexcitation, its dynamics is inherently ultrafast, which is in fact observed recently [19].

When charged particles run in space, they radiate electromagnetic (EM) waves. Thus a pulsed photocurrent excited by sub-picosecond lasers leads to the generation of EM waves in terahertz (THz) frequency range [20]; nowadays this phenomenon is widely utilized for THz light sources by comprising photoconductive switches under the electric bias. Conversely, if we analyze the temporal/spectral dynamics of the emitted THz waves, it is possible to elucidate the dynamics of charges in a crystal with sub-picosecond time resolution, which cannot be attained by using conventional electric circuits with electrodes. Such THz emission spectroscopy has become a powerful tool to study the ultrafast spontaneous photocurrent, as exemplified in several semiconductors and quantum dots [21-24]. There are known to be several EM-wave sources in solids under pulsed photoexcitation, such as optical rectification (OR) [23] and coherent phonons [25,26]. The shift current corresponds to one of the OR processes with the interband photoexcitation, while we refer the other conventional OR with the intraband (below-band-gap) excitation to the "intraband OR" hereafter [6].

In this study, the observed THz spectra are analyzed in detail by using the procedure of a factor analysis, enabling us to resolve electron dynamics of different origins. The excitation spectrum nicely follows that predicted by the first-principles calculations. It is concluded that the shift current dominates the photocurrent in SbSI when excited above the band gap, with the averaged charge shift of ~0.18 Å in the Sb-S and Sb-I bonds, which relaxes with the time-constant of 0.4~0.5 ps after transferring a portion of charges to the neighbouring sites.

## II. EXPERIMENTAL RESULTS

SbSI is a prototypical ferroelectric semiconductor with a band gap of 1.9-2.0 eV and spontaneous polarization of ~25 µC/cm$^2$ [27,28], proposed to be a candidate for the shift current solar cell [29]. The ferroelectric polarization appears along the *c*-axis of the crystal (Fig. 1(a)) [30] near room temperature, along which quasi one-dimensional structures constitute. Photovoltaic properties of SbSI have been studied intensively [28,31,32]. We measured the THz emission from SbSI single crystals in reflection geometry using variable wavelength laser (130 fs, 1 kHz, spot size ~90 µm in diameter) for excitation. A ZnTe (110) crystal (500 µm thickness) was used for the electro-optic (EO) sampling to analyse the THz waveform. The sample was mounted on a Peltier stage and poled before each measurement by cooling through the $T_C$ under the external electric field. A reference ZnTe crystal was mounted next to the sample to evaluate the time-zero and instrumental function of the THz waveform measurements.

Figure 1(b) shows the representative THz waveforms emitted from an SbSI single crystal upon pulsed photoexcitation. When the bulk polarization is reversed by an electric poling procedure, the phase of the THz wave is perfectly reversed, indicating the opposite current flow in the sample. The THz wave intensity is found to be comparable to that in a GaAs(110) crystal measured under the same setup. This THz emission can be seen only below the Curie temperature ($T_C$ = 295 K) [28], and the temperature variation of its absolute amplitude ($\sqrt{I_{THz}}$; square root of the integrated power spectrum) nearly follows that of the ferroelectric polarization (Figs. 1(c) and 1(d)).

To uncover the photocarrier dynamics with and without Berry phase contributions, here we perform terahertz emission spectroscopy by sweeping the excitation photon energy across the band gap. Figure 2(a) compares the THz waveforms excited at 2.1 eV and 1.8 eV, just above and below the band gap energy of SbSI (interband and intraband excitation, respectively). It is seen that the interband excitation causes a delay of the THz peak (~220 fs) from the time zero, which is accompanied by a prolonged dip in the end of the emission (extending to ~1 ps, indicated by an arrow). The presence or absence of the delay in the THz peak appears due to the difference in the emission mechanism: the shift current follows a real transition of the electron cloud, whereas the intraband excitation drives only virtual excitation, i.e., deformation of the wave function in the valence band.

By using the action spectra measured by sweeping the excitation photon energy (22 points) from 0.5 to 2.6 eV, we performed a factor analysis [33]. We employed this technique to single out a THz waveform from possibly mixed signals of multiple origins. This technique has been conventionally utilized to separate base waveforms from the fluorescence action spectra [34]. We found that all the spectra can be expressed by two base waveforms shown in Fig. 2(b), which can be attributed to the cases of inter- and intraband photoexcitation (Fig. 2(a)). The corresponding THz amplitude spectra, proportional to the photocurrent density, are shown in Fig. 2(c), by taking the effective generation length into account. It is seen that the shift current emerges only above the band gap, with a peak and a shoulder around 2.10 eV and 2.35 eV, respectively, while the intraband OR exists constantly within the band gap energy. The reduction of the intraband OR above the band gap is due mainly to the sudden decrease in the penetration depth of the incident light, which limits the effective volume generating the THz waves. Considering these factors, the photocurrent generated by the shift mechanism is found to be ~100 times larger in density.

The experimental shift current spectrum nicely follows that deduced from the first-principles calculation with the Berry connection integrated [35], although the fine structures above 2.2 eV are somewhat smeared out. Thus, we can ascribe the THz wave emission for the interband excitation to the shift current, whereas for the intraband one to the conventional intraband OR. We will further discuss the reasoning for this assignment in the next section.

A transient current can be generated through a virtual intraband optical excitation in polar crystals [6]. The induced nonlinear polarization follows the time profile of the excitation laser pulse (Fig. 3(a)), whose time-derivative leads to the transient charge current, but not to the net flow of current, *i.e.*, its integral over time vanishes. This current, the intraband OR, can be a source of the THz emission, with the waveform following the second derivative of the nonlinear polarization. By the shift-current excitation, in contrast, the electron cloud transits in real space along the chemical bonds on a primary-process time scale. Subsequently this charge shift relaxes in the absence of optical field, in many cases via scattering by phonons. Most importantly, a portion of charges is left in the neighbouring site upon this relaxation, leading to a net current flow which survives even after the integration over time.

The respective carrier dynamics discussed above can be experimentally visualized by integrating the THz waveforms. Experimentally obtained carrier dynamics was fitted by

$$J(t) = \frac{1}{\sqrt{2\pi}\tau_r} \exp\left(\frac{-t^2}{2\tau_r^2}\right) \otimes \left[J_{\text{shift}}\delta(t) - J_{\text{decay}}u(t)\exp\left(-\frac{t}{\tau}\right)\right], (3)$$

where the first part represents the incident laser pulse ($\tau_r = 55$ fs for the case of our 130 fs incident Gaussian laser pulse), $\delta(t)$ the delta function, $u(t)$ the step function, and $\tau$ the scattering (relaxation) time [19]. Figure 3(b) illustrates the retrieved current dynamics in the time domain. The intraband OR shows the waveform to be a derivative of the incident laser pulse (approximately of Gaussian form). In stark contrast, the shift current appears with some delay and with a slow relaxation (swing over), and carries the net current. The decay (scattering) time $\tau$ was found to be 0.4~0.5 ps for SbSI by following eq. (3). It is reasonable to ascribe the decay to the scattering by the phonons in SbSI, some of which have the frequency around 1~3 THz [36, 37]. This decay time is found to be nearly independent of temperature and incident photon energy except for the cases near the $T_C$ and around the band gap energy (Figs. 3(c) and 3(d)). The prolonged $\tau$ near the $T_C$ and at the band gap energy is related to the dielectric screening and to the absence of excess energy to provide thermalizing phonons at the band-edge [38].

By using the THz signals from the reference ZnTe crystal with known nonlinear optical constants [39], the shift distance i.e., the spatial displacement of the electron cloud, was evaluated to be ~0.18 Å from the data shown in Fig. 3(b). This value is comparable to the previously reported ones in CdSe [22] and $Bi_2Se_3$ [19], considering the uncertainty in the estimation (factor of 2).

For the generation of shift current, the characteristic incident power dependence is predicted [8]. Including higher-order corrections, the shift current can be expressed as

$$j = \frac{\pi e^2 |E_{exc}|^2}{\hbar^2 \omega_{exc}^2} \int \frac{d^3k}{(2\pi)^3} |v_{12}|^2 \frac{\Gamma}{\sqrt{4\left(\frac{e|E_{exc}||v_{12}|}{\hbar \omega_{exc}}\right)^2 + \Gamma^2}} R_k \, \delta(\epsilon_1 - \epsilon_2 + \hbar \omega_{exc}). \quad (4)$$

Here, $\Gamma = 2\pi/\tau$ is the scattering rate, and $\epsilon_1$ ($\epsilon_2$) the energy of the valence (conduction) band [8]. Thus the shift current shows specific incident power ($|E_{exc}|^2$) dependence. Figure 4(a) displays the experimental incident power dependence. The fitting to eq. (4) yields $v_{12}$ to be ~$2.3 \times 10^4$ m/s. Figure 4(b) represents the optical conductivity spectrum $\sigma(\omega)$ of SbSI at 282 K, which is derived by Kramers-Kronig transformation of the reflectance. We used a multiple dipole model to analyze the spectrum [40]. The transition matrix element at 2.3 eV averaged by three peaks, $v_{mean}$, is estimated to be $2.0 \times 10^4$ m/s. This is very close to the value evaluated above from the incident power dependence.

Lastly we demonstrate the incident polarization dependence. Figure 5 compares the polarimetry of zero-bias photocurrent evaluated by using a conventional electric circuit [32] and that of the THz emission. The anisotropy will be discussed by following the motion of electron cloud on the chemical bonds (Fig. 5(a)). With the interband excitation at photon energy ($\hbar \omega_{exc}$) of 2.3 eV (Fig. 5(b)), a finite shift current along the *c*-axis runs into the electrodes. Both the zero-bias photocurrent and THz emission show sinusoidal polarization dependence with a maximum near 90 degree (Figs. 5(b) and 5(d)), which is consistent with the previous reports [28]. In contrast, the zero-bias photocurrent is absent for the intraband excitation with $\hbar\omega_{\text{exc}} = 1.0$ eV (Fig. 5(c)). The THz emission for this case shows distinct polarization dependence with a maximum near 0 degree. The THz emission perpendicular to the *c*-axis ($E(\omega_{\text{THz}}) \perp c$) is found to be small for both cases of excitation. The SbSI belongs to the $C_{2v}$ (*mm2*) point group in the ferroelectric phase, with the photo-conductivity ($\sigma^{(2)}_{ijk}$) tensors of the form

$$\sigma^{(2)}_{ijk} = \begin{pmatrix} 0 & 0 & 0 & 0 & \sigma^{(2)}_{aca} & 0 \\ 0 & 0 & 0 & \sigma^{(2)}_{bbc} & 0 & 0 \\ \sigma^{(2)}_{caa} & \sigma^{(2)}_{cbb} & \sigma^{(2)}_{ccc} & 0 & 0 & 0 \end{pmatrix}. \quad (5)$$

The nonlinear optical tensor ($\chi^{(2)}_{ijk}$) has the same symmetry. By fitting (Figs. 5(d) and 5(e)), we can estimate the effective tensor elements to be $\chi^{(2)}_{caa}/\chi^{(2)}_{ccc} = 0.14$, $\chi^{(2)}_{aca}/\chi^{(2)}_{ccc} = 0.07$ for the intraband excitation (intraband OR; $\hbar\omega_{\text{exc}} = 1.0$ eV), and $\sigma^{(2)}_{aca}/\sigma^{(2)}_{ccc} = 0.19$, $\sigma^{(2)}_{caa}/\sigma^{(2)}_{ccc} = 1.75$ for the interband excitation (shift current; $\hbar\omega_{\text{exc}} = 2.3$ eV). Here each tensor element includes Fresnel factors, and we set $a = b$ for simplicity. We speculate that the observed enhancement of $\sigma^{(2)}_{caa}$ ($= \sigma^{(2)}_{cbb}$) in the shift current is related to the electron transition between S and I sites and that between S and Sb sites (Fig. 5(a)) [41].

**III. Conclusion**

The original expression of shift current (eq. (1)) treats solely the excitation process, while later it has been extended to incorporate the broadening of the electronic bands (i.e., damping/scatterings) [8]. The finite photocurrent generated by the shift mechanism is affected by both the excitation and relaxation processes, as verified in the present experiment. By analyzing the terahertz electromagnetic spectra radiated from the photocurrent, we have uncovered the ultrafast dynamics of the shift current including its relaxation. The intriguing characteristics of the shift current, reflecting the Berry phase contributions in the electronic process, were experimentally traced with photoexcitation under varying temperature, poling, excitation photon energy, and polarization conditions. It is demonstrated that the sub-picosecond photocurrent shows a distinct transition from the intraband optical rectification to the shift excitation in their time-profile when the excitation photon energy crossing the band gap; only the latter carries a net current to the electrodes. The shift current spectrum is found to be consistent with the band calculation incorporating the Berry connection. We also discussed the distance and direction of the shift in the electron cloud, which gives the impulse excitation of phonons in the corresponding chemical bonds. These phonons finally scatter the electron in its relaxation process in a sub-picosecond time scale, and help transfer the portion of electrons in the neighboring sites. Understanding these features of the shift current would materialize novel dissipation-less ultrafast electronic devices.

## Acknowledgments

We thank D. Maryenko, and W. Koshibae for stimulating discussions. This research is supported by JSPS KAKENHI Grant Numbers 24224009, 16H00981, 16K13705, and 17H02914. M.N. J.F. and N.O. are supported by PRESTO, JST (JPMJPR16R5, JPMJPR15R5, JPMJPR17I3, respectively). YZ is supported by German Research Foundation (DFG) SFB 1143.

**Figures**

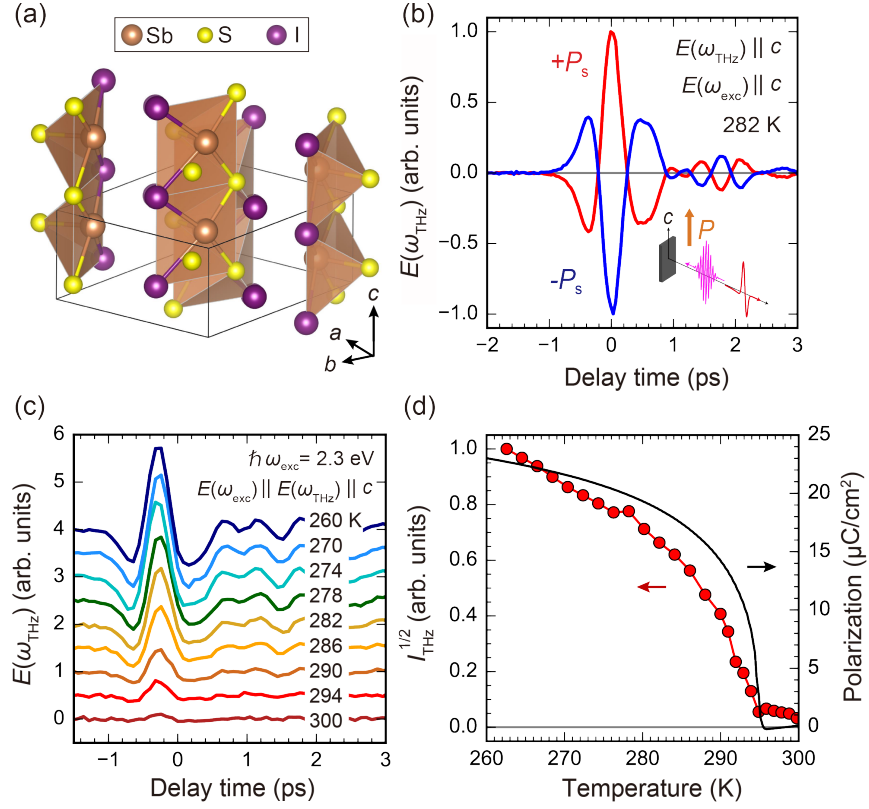

FIG. 1. Emission of terahertz (THz) waves from SbSI. (a) Crystal structure of SbSI. (b) THz waves emitted from SbSI in the ferroelectric phase after ±$P$ poling (±2.0 kV/cm, respectively). Excitation photon energy ($\hbar\omega_{\text{exc}}$) is 2.3 eV (400 nJ). (c) Temperature dependence of the THz waveform. (d) Temperature dependence of the THz intensity, plotted together with the pyroelectric polarization. $I_{\text{THz}}$ stands for the frequency integration of the power spectrum in the Fourier space, which is proportional to the square of the source current.

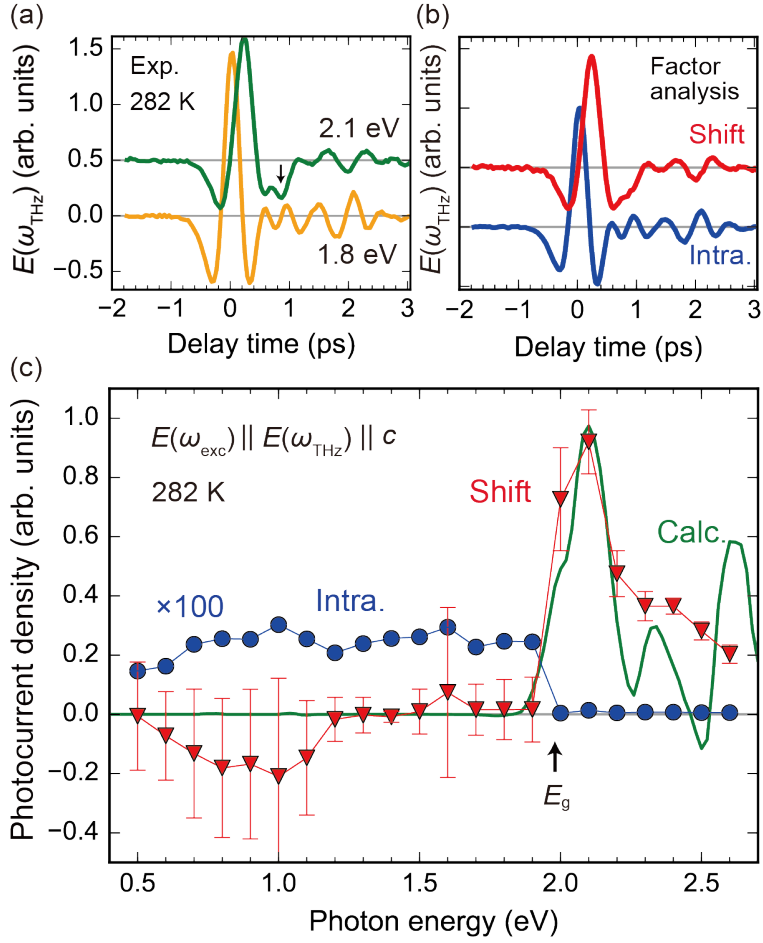

FIG. 2. Action spectra of the shift current deduced by the factor analysis. (a) Experimental THz waveforms for the excitation above and below the band-gap ($E(\omega_{\text{THz}}) \parallel c$, 400 nJ, offset for clarity). (b) Extracted base waveforms by using the factor analysis (covariance matrix calculation) from 22 data sets. (c) Action spectra of shift current (Shift) and intraband (Intra.) optical rectification (OR). Error bars reflect the unique variance in the factor analysis, which are smaller than the marker for the case of intraband OR. Result of the first-principles calculation (normalized amplitude of the shift current) is also shown.

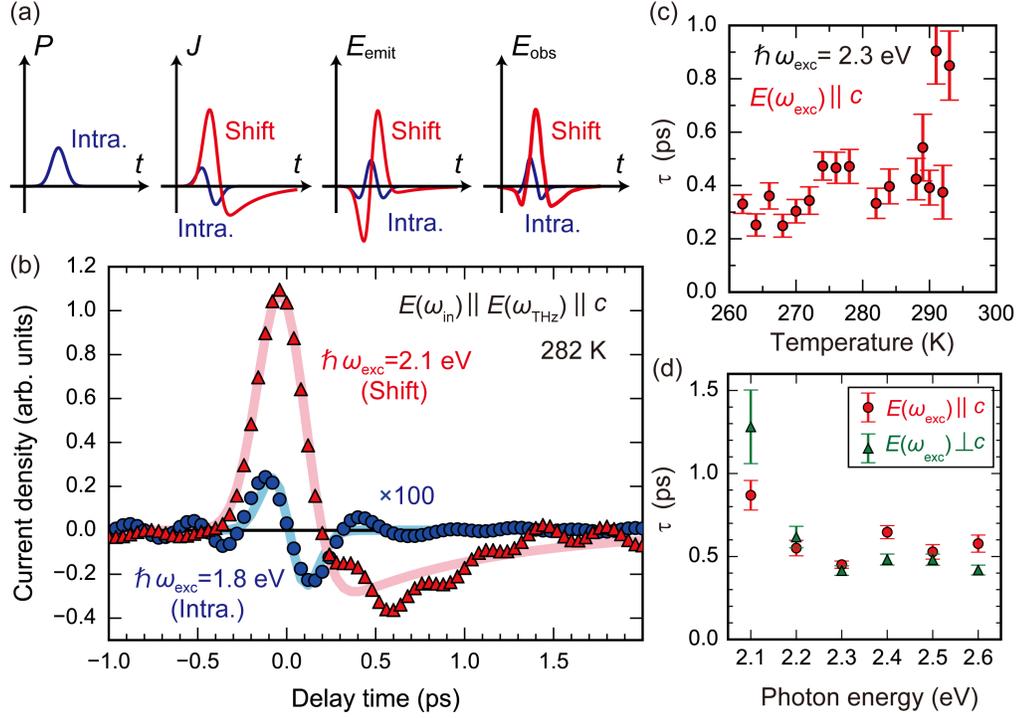

FIG. 3. Transient nonlinear polarization and carrier dynamics. (a) Schematics of the nonlinear polarization ($P$), transient current ($J$), and emitted electric field ($E_{\text{emit}}$) for the shift current (Shift) and intraband (Intra.) photoexcitation. The detected THz electric fields are further modified ($E_{\text{obs}}$) due to the diffraction and instrumental factors. (b) Retrieved current dynamics for the shift current and intraband photoexcitation. Solid lines represent the fitting curves. Shift current accompanies a swing over (relaxation) of the charge with a decay time of ~0.5 ps, whereas the intraband photoexcitation appears only within the incident pulse duration. (c),(d) Temperature and incident photon energy dependence of the decay time $\tau$.

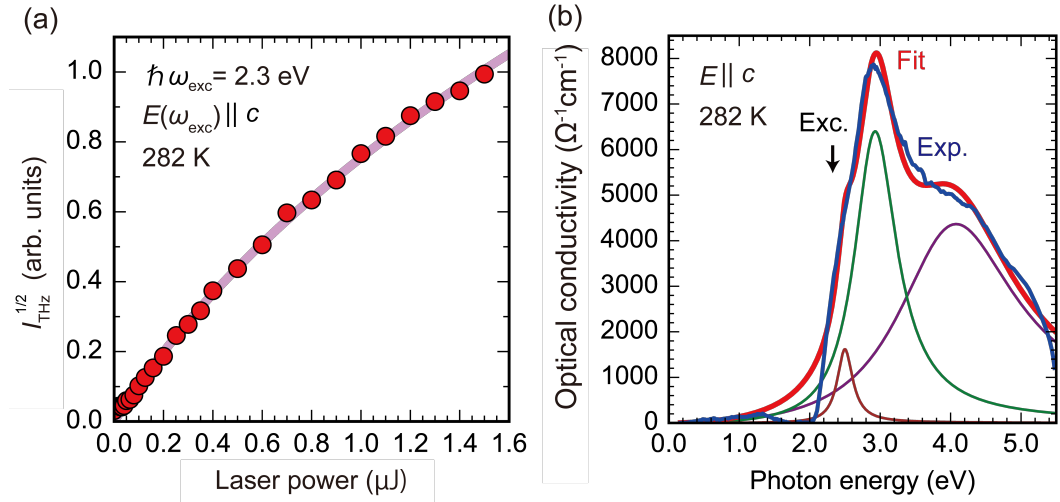

FIG. 4. Incident laser power dependence of THz emission, and optical conductivity spectrum. (a) Amplitude of the emitted THz ($I_{THz}^{1/2}$) showing sublinear incident power dependence. Solid line shows a fit by eq. (4). (b) Fitting analysis on the optical conductivity spectrum.

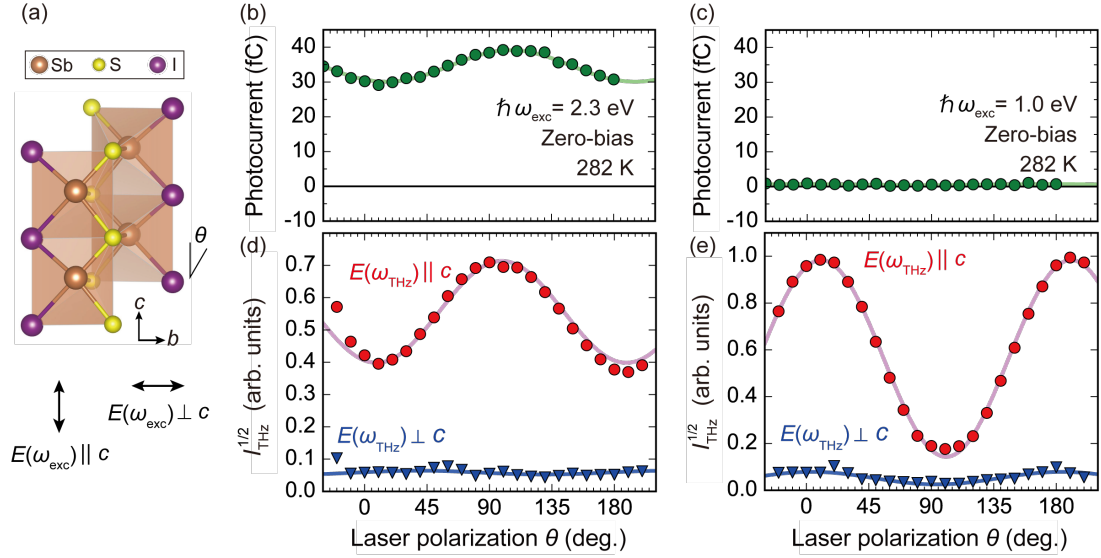

FIG. 5. Polarimetry of shift current and intraband optical rectification. (a) Schematic illustration of the connected polyhedra and definition of angle $\theta$. (b),(c) Incident polarization dependence of the pulsed zero-bias photocurrent with excitation photon energy ($\hbar\omega_{exc}$) of 2.3 and 1.0 eV (above and below the band gap) measured by a transimpedance amplifier through the electrodes. The zero degree corresponds to the polarization parallel to the $c$-axis. (d),(e) Corresponding amplitude spectra of the emitted THz waves ($\sqrt{I_{THz}}$). Solid lines represent the fittings considering the symmetry-allowed tensor elements (see eq. (5) in the text).

# References

[1] R. Resta, D. Vanderbilt, *Topics in Applied Physics* **105** (Springer, Berlin, Heidelberg, 2007).

[2] R. Resta, *J. Phys.: Condens. Matter* **22**, 123201 (2010).

[3] S. M. Young, and A. M. Rappe, *Phys. Rev. Lett.* **109**, 116601 (2012).

[4] R. von Baltz, and W. Kraut, *Phys. Rev. B* **23**, 5590-5596 (1981).

[5] B. I. Sturman, and V. M. Fridkin, *The Photovoltaic and Photorefractive Effects in Noncentrosymmetric Materials*. (Gordon and Breach Science Publishers, Philadelphia, 1992).

[6] J. E. Sipe, and A. I. Shkrebtii, *Phys. Rev. B* **61**, 5337-5352 (2000).

[7] P. Král, *J. Phys.: Condens. Matter* **12**, 4851-4868 (2000).

[8] T. Morimoto, and N. Nagaosa, *Sci. Adv.* **2**, e1501524 (2016).

[9] T. Morimoto, and N. Nagaosa, *Phys. Rev. B* **94**, 035117 (2016).

[10] K. W. Kim, T. Morimoto, and N. Nagaosa, *Phys. Rev. B* **95**, 035134 (2017).

[11] W. T. H. Koch, R. Munser, W. Ruppel, and P. Würfel, *Ferroelectrics* **13**, 305-307 (1976).

[12] S. Y. Yang, L. W. Martin, S. J. Byrnes, T. E. Conry, S. R. Basu, D. Paran, L. Reichertz, J. Ihlefeld, C. Adamo, A. Melville, Y.-H. Chu, C.-H. Yang, J. L. Musfeldt, D. G. Scholom, J. W. Ager III, and R. Ramesh, *Appl. Phys. Lett.* **95**, 062909 (2009).

[13] M. Nakamura, F. Kagawa, T. Tanigaki, H. S. Park, T. Matsuda, D. Shindo, Y. Tokura, and M. Kawasaki, *Phys. Rev. Lett.* **116**, 156801 (2016).

[14] K. Kushnir, M. Wang, D. Fitzgerald, K. J. Koski, and L. V. Titova, *ACS Energy Letters* **2**, 1429 (2017).

[15] M. Nakamura, S. Horiuchi, F. Kagawa, N. Ogawa, T. Kurumaji, Y. Tokura, and M. Kawasaki, *Nature Commun.* **8**, 281 (2017).

[16] L. Z. Tan, F. Zheng, S. M. Young, F. Wang, and A. M. Rappe, *npj Comp. Mater.* **2**, 16026 (2016).

[17] A. M. Cook, B. M. Fregoso, F. de Juan, S. Coh, and J. E. Moore, *Nature Commun.* **8**, 14176 (2017).

[18] K. T. Butler, J. M. Frost, and A. Walsh, *Energy Environ. Sci.* **8**, 838-848 (2015).

[19] L. Braun, G. Mussler, A. Hruban, M. Konczykowski, T. Schumann, M. Wolf, M. Münzenberg, L. Perfetti, and T. Kampfrath, *Nature Commun.* **7,** 13259 (2016).

[20] X.-C. Zhang, B. B. Hu, and D. H. Auston, *Appl. Phys. Lett.* **56**, 1011-1013 (1990).

[21] J. H. Son, T. B. Norris, and J. F. Whitaker, *JOSA B* **11**, 2519-2527 (1994).

[22] N. Laman, M. Bieler, and H. M. van Driel, *J. Appl. Phys.* **98**, 103507 (2005).

[23] Z. Mics, H. Nemec. I. Rychetsky, P. Kuzel, P. Formanek, P. Maly, and P. Nemec, *Phys. Rev. B* **83**, 155326 (2011).

[24] X.-C. Zhang, Y. Jin, and X. F. Ma, *Appl. Phys. Lett.* **61**, 2764-2766 (1992).

[25] T. Dekorsy, H. Auer, H. J. Bakker, H. G. Roskos, and H. Kurz, *Phys. Rev. B* **53**, 4005-4014 (1996).

[26] M. Sotome, N. Kida, R. Takeda, and H. Okamoto, *Phys. Rev. A* **90**, 033842 (2014).

[27] E. Fatuzzo, G. Harbeke, W. J. Merz, R. Nitsche, H. Roetschi, and W. Ruppel, *Phys. Rev.* **127**, 2036-2017 (1962).

[28] V. M. Fridkin, *Photoferroelectrics* (Springer-Verlag, New York 1979).

[29] K. T. Butler, S. McKechnie, P. Azarhoosh, M. van Schilfgaarde, D. O. Scanlon, and A. Walsh, *Appl. Phys. Lett.* **108**, 112103 (2016).

[30] K. Momma, and F. Izumi, *J. Appl. Crystallogr.* **44**, 1272-1276 (2011).

[31] V. M. Fridkin, and A. I. Rodin, *Physica Status Solidi* (a) **61**, 123-126 (1980).

[32] N. Ogawa, M. Sotome, Y. Kaneko, M. Ogino, and Y. Tokura, *Phys. Rev. B* **96**, 241203 (2017).

[33] C. M. Bishop, *Pattern Recognition and Machine Learning* (Springer-Verlag New York, 2006).

[34] C. A. Stedmon, and R. Bro, *Limnology and Oceanography*: *Methods* **6**, 572-579 (2008).

[35] In the calculation of the shift current spectrum, the band structure of SbSI was calculated based on the density-functional theory (DFT) with the localized atomic orbital basis and the full potential as implemented in the full-potential local-orbital (FPLO) code (K. Koepernik and H. Eschrig, *Phys. Rev. B* **59,** 1743-1757 (1999)). The exchange-correlation functions were considered at the generalized gradient approximation (GGA) level (J. P. Perdew, K. Burke, and M. Ernzerhof, *Phys. Rev. Lett.* **77,** 3865 (1996)). We use the experimentally reported lattice structure. By projecting the wave function in atomic basis into a reduced symmetric atomic orbital like Wannier functions (Sb-$p$, S-$p$ and I-$s$, $p$ orbitals), we constructed tight-binding Hamiltonians with 96 bands and computed the shift current response. The obtained shift current spectrum was shifted in energy (~0.09 eV) due to the underestimation of the band gap, convoluted with the experimental incident laser spectrum, and plotted in Fig. 2(c).

[36] D. K. Agrawal, and C. H. Perry, *Phys. Rev. B* **4**, 1893-1902 (1971).

[37] H. Buhay, and C. H. Perry, *J. Phys. Chem.* **80**, 1208-1211 (1976).

[38] I. V. Bazarov, B. M. Dunham, X. Liu, M. Virgo, A. M. Dabiran, F. Hannon, and H. Sayed, *J. Appl. Phys.* **105**, 083715 (2009).

[39] Q. Wu, and X.-C. Zhang, *Appl. Phys. Lett.* **68**, 1604-1606 (1996).

[40] M. S. Dresselhaus. *Optical properties of solids. Proceedings of the International School of Physics* (Academic Press. NY, 1966).

[41] D. Amoroso, and S. Picozzi, *Phys. Rev. B* **93**, 214106 (2016).